\documentclass[11pt, a4paper]{article}
\usepackage[]{graphicx}
\usepackage[]{color}
\makeatletter
\def\maxwidth{ %
  \ifdim\Gin@nat@width>\linewidth
    \linewidth
  \else
    \Gin@nat@width
  \fi
}
\makeatother

\definecolor{fgcolor}{rgb}{0.345, 0.345, 0.345}

\usepackage{framed}
\makeatletter
 {\par\unskip\endMakeFramed%
 \at@end@of@kframe}
\makeatother

\definecolor{shadecolor}{rgb}{.97, .97, .97}
\definecolor{messagecolor}{rgb}{0, 0, 0}
\definecolor{warningcolor}{rgb}{1, 0, 1}
\definecolor{errorcolor}{rgb}{1, 0, 0}

\usepackage{alltt}

\usepackage{verbatim}
\usepackage[vmargin=3cm, hmargin=2cm]{geometry}
\usepackage{url}
\usepackage{hyperref}
\usepackage{fancyhdr}
\usepackage{color}
\usepackage{amsmath, amssymb}
\usepackage{longtable}
\usepackage{lscape}
\usepackage[super, comma,numbers]{natbib}
\usepackage{xspace}
\usepackage{booktabs}
\usepackage{newfloat}
\usepackage{rotating}
\usepackage{scrextend}
\usepackage{csquotes}
\DeclareFloatingEnvironment[
    placement=!htbp
]{listing}
\usepackage{fancyvrb}
\usepackage{lmodern}
\usepackage{nameref}
\usepackage[T1]{fontenc}
\usepackage{listings}
\usepackage{setspace}
\usepackage[font=sf, labelfont={sf}, margin=1cm]{caption}

\usepackage[latin1]{inputenc}

\IfFileExists{upquote.sty}{\usepackage{upquote}}{}
\begin{document}
\begin{center}
    \Large
    \textbf{A Bayesian time-to-event pharmacokinetic model for sequential phase I dose-escalation trials with multiple schedules}

    \vspace{0.4cm}
    \large
    \textbf{Burak K\"ursad G\"unhan},\footnote{\textit{Department of Medical Statistics, University Medical Center G\"ottingen, G\"ottingen, Germany} \label{goettingen}} \footnote{\textit{Correspondence to: Burak K\"ursad G\"unhan; email: \texttt{burak.gunhan@med.uni-goettingen.de}}} \textbf{Sebastian Weber},\footnote{\textit{Novartis Pharma AG, Basel, Switzerland} \label{novartis}} \textbf{Abdelkader Seroutou},\footref{novartis}  \textbf{Tim Friede}\footref{goettingen} 
    \vspace{0.9cm}
\end{center}

Phase I dose-escalation trials constitute the first step in investigating the safety of potentially promising drugs in humans. Conventional methods for phase I dose-escalation trials are based on a single treatment schedule only. More recently, however, multiple schedules are more frequently investigated in the same trial. Here, we consider sequential phase I trials, where the trial proceeds with a new schedule (e.g. daily or weekly dosing) once the dose escalation with another schedule has been completed. The aim is to utilize the information from both the completed and the ongoing dose-escalation trial to inform decisions on the dose level for the next dose cohort. For this purpose, we adapted the time-to-event pharmacokinetics (TITE-PK) model, which were originally developed for simultaneous investigation of multiple schedules. TITE-PK integrates information from multiple schedules using a pharmacokinetics (PK) model. In a simulation study, the developed appraoch is compared to the bridging continual reassessment method and the Bayesian logistic regression model using a meta-analytic-prior. TITE-PK results in better performance than comparators in terms of recommending acceptable dose and avoiding overly toxic doses for sequential phase I trials in most of the scenarios considered. Furthermore, better performance of TITE-PK is achieved while requiring similar number of patients in the simulated trials. For the scenarios involving one schedule, TITE-PK displays similar performance with alternatives in terms of acceptable dose recommendations. The \texttt{R} and \texttt{Stan} code for the implementation of an illustrative sequential phase I trial example is publicly available (\href{https://github.com/gunhanb/TITEPK_sequential}{https://github.com/gunhanb/TITEPK\_sequential}). In sequential phase I dose-escalation trials, the use of all relevant information is of great importance. For these trials, the adapted TITE-PK which combines information using PK principles is recommended.\\


\section{Background}  \label{sec:1}
Phase I dose-escalation trials constitute the first step in investigating the safety of potentially promising drugs in humans \cite{Tourneau2009}. In oncology, such trials focus on identifying the maximum tolerated dose (MTD) through a series of dose-escalation steps. Dose-escalation trials traditionally enroll small cohorts of patients who are treated in
cycles. Typically, the estimation of the MTD is based on the toxicity data of the first cycle only. The observed toxicities are classified into dose-limiting toxicities (DLT) and non-DLT. Each time a cohort completes the first cycle at a given dose level, the available data are assessed to decide how the trial proceeds. A commonly accepted target for the MTD in oncology is to allow for a DLT probability of 33\% per cycle of treatment.

Standard statistical methods include adaptive model based approaches such as the continual reassessment method (CRM) \cite{10.2307/2531628} or the Bayesian logistic regression model (BLRM) \cite{SIM:SIM3230}. The BLRM is
a two-parameter version of the CRM which utilizes the
escalation with the overdose control (EWOC) \cite{SIM:SIM793} criterion. The EWOC criterion aims to reduce the risk of overdosing patients by choosing doses with a posterior probability of being above the true MTD lower than a feasibility bound.

In addition to the dose administered, the frequency of administration, known as the schedule, is a crucial part of a treatment plan of any phase I trial. In practice, sometimes it is required to investigate multiple schedules, e.g. a dose given once a day or an adequately larger dose given once a week. Hence, the probability of DLT for each patient is a function of both the dose and the schedule. Simultaneous investigation of dose and schedule within a phase I trial has gained some attention in the literature. In such trials, the doses and the schedules are altered for different cohorts of patients within the same trial. Methods for simultaneous investigation of dose and schedule combination include a Bayesian time-to-event model by Braun et al \cite{Braun2007} and the partial order continual reassessment method by Wages et al \cite{Wages2014}. Recently, G\"unhan et al \cite{titePK} proposed an alternative dose-schedule finding method, a Bayesian time-to-event pharmacokinetics model (TITE-PK), which uses pharmacokinetics (PK) principles. Unlike other phase I methods, TITE-PK makes use of an exposure-response model that is often more informative than a standard dose-response model. TITE-PK models the relationship between time-to-first DLT and an exposure measure of the drug obtained by a \emph{pseudo}-PK model in a Bayesian model-based approach. TITE-PK has been shown to have desirable operating characteristics in terms of finding an acceptable dose and schedule simultaneously in simulation studies \cite{titePK}. 

In this paper, we consider an alternative phase I design in which multiple treatment schedules are investigated sequentially, rather than simultaneously.  The schedules are denoted by $S_{i}$ where $i=1, 2, \ldots, k$. The sequential multiple schedule design proceeds as follows. In the first step, cohorts of patients are enrolled with $S_{1}$ and the trial is continued until the MTD is declared for $S_{1}$. In the second step, the trial continues with schedule $S_{2}$ and the starting dose can be informed from the $S_{1}$. Dose-escalation decisions are informed by utilizing information from both schedules $S_{1}$ and $S_{2}$. That is, data from both the completed schedule $S_{1}$ and the ongoing schedule $S_{2}$ are integrated. Once the MTD for the Schedule $S_{2}$ is determined, the trial can continue with schedule $S_{3}$ and so on.

A sequential phase I trial with different strata, where strata may correspond to different patient populations, formulations, or treatment schedules etc., also called as a bridging trial, was considered by Liu et al \cite{Liu2015} among others \cite{doi:10.1080/19466315.2016.1174149, Ollier2019, li2020}. Liu et al \cite {Liu2015} introduced the bridging CRM to borrow information from different strata. B-CRM takes into account potential heterogeneity between different strata using a Bayesian model averaging approach. Neuenschwander et al \cite{doi:10.1080/19466315.2016.1174149} suggest the use of BLRM with a meta-analytic-predictive (MAP) prior \cite{BIOM:BIOM12242} approach (BLRM-MAP) to take advantage of the completed trial with different strata.

Borrowing approaches are based on discounting the existing information at the cost of increasing the needed saample size to achieve an acceptable performance in a new trial. Here we suggest the use of a modelling approach based on PK principles in order to increase the statistical efficiency. Therefore, we adapted the TITE-PK to design and analyze sequential phase I trials with multiple schedules. In the first step, TITE-PK is used to inform dose-escalation decisions for schedule $S_{1}$ until the MTD is declared or the trial is stopped. In the next step, TITE-PK models the data from both the completed ($S_{1}$) and the ongoing ($S_{2}$) trial directly, but only recommending doses for Schedule $S_{2}$. TITE-PK can be used for any number of schedules. We investigate the operating characteristics of TITE-PK for phase I trials with one schedule and sequential phase I trials with multiple schedules through simulations. We provide simulation results comparing the performance of TITE-PK to CRM and BLRM for phase I trials involving one schedule and to B-CRM and BLRM-MAP for sequential phase I trials involving multiple schedules. 

This paper is organized as follows. In the following section, we describe an illustrative phase I trial example from oncology which
investigated daily and weekly treatment schedules. Then, we
describe the adapted TITE-PK for sequential investigation of multiple schedules. The
performance of TITE-PK and comparators are studied in simulations. Later, different methods are applied to the illustrative example. We close with discussion and conclusions.

\subsection{Illustrative example: Everolimus trial}

Everolimus (RAD001) is an oral inhibitor of mammalian target of 
rapamycin, that has been developed as an antitumor agent \cite{EVEROLIMUS}.
Everolimus is approved by the US FDA to treat various conditions 
including certain types of pancreatic cancer and gastrointestinal 
cancer \cite{EVEROLIMUS} and certain type of tuberous sclerosis \cite{RAD2}. The elimination half-life and the absorption 
rate of everolimus for cancer patients were reported as 30 (hours) and 2.5 (1/hours),
respectively \cite{doi:10.1200/JCO.2007.14.0988}.
Everolimus was included in a phase Ib trial in combination with 
standard of care (etoposide and cisplatin chemotherapy) to identify 
a feasible dose and schedule in the treatment of small cell lung cancer 
(ClinicalTrials.gov identifier: NCT00466466) \cite{doi:10.1093/annonc/mdt535}.
The trial was open-label and multi-centered. Patients were assigned 
alternately to either weekly or daily schedules of everolimus in 
treatment cycles of 21 days. In the everolimus trial, doses in both schedules were escalated simultaneously and analysed separately from one another. A Bayesian time-to-event model \cite{Cheung2000TITECRM} was used to inform the dose-escalation decisions. The final data can be obtained from 
the supplementary material of Besse et al \cite{doi:10.1093/annonc/mdt535}. The dataset is displayed in Table~\ref{t01:RawDataRAD}. All DLT 
were reported at day 15.
Based on investigator and medical monitor opinion, 2.5 
mg/$\text{m}^2$ with daily schedule was identified as the MTD \cite{doi:10.1093/annonc/mdt535}.

\begin{table}
\centering
\caption{Data of the everolimus trial. The treatment schedules which are used, the doses which are administered in mg/$\text{m}^2$, number of patients, and number of DLT are given.}
\label{t01:RawDataRAD}
\begin{tabular}{llll}
  \toprule
Schedule & Dose & Number      & Number \\ 
        & (mg/$\text{m}^2$) & of patients & of DLT   \\ 

  \midrule
Weekly & 20.0 & 5  & 0 \\ 
Weekly & 30.0 & 13 & 4 \\ 
Daily  & 2.5  & 4  & 2 \\ 
Daily  & 5.0  & 6  & 3 \\ 
   \bottomrule
\end{tabular}
\end{table}

We used this trial to illustrate the TITE-PK approach for sequential designs, because 
(1) the trial evaluated two different schedules (weekly and daily dosing) and (2) the large number of DLT allows a good assessment on the performance of the TITE-PK. We will analyse the final dataset
as if the trial had been conducted sequentially, specifically assuming $S_{1}$ is weekly schedule 
and $S_{2}$ is daily schedule.

\section{Methods} 
\subsection{TITE-PK for sequential phase I trials} 
TITE-PK for simultaneous investigation of multiple schedules in phase I trials were introduced in G\"unhan et al \cite{titePK}, here we adapt it for sequential investigation of multiple schedules. The time-to-first DLT events are modeled using a time-varying (non-homogeneous) Poisson process. The hazard function is assumed to depend on an exposure measure of the drug (E($t$)):
\begin{align}
  h(t) = \beta \, E(t) \label{eq:exposure}
\end{align}
where $\beta$ is the only parameter to estimate in the model.

The exposure measure is calculated using a pseudo-PK model which consists of two ordinary differential equations:
\begin{align}
  &\frac{dC(t)}{dt} = - k_{e} \, C(t)\,\,\,\, \text{and} \,\,\,\, C(0) = 0 \nonumber \\
  &\frac{d C_{\text{eff}}(t)}{dt} = k_{\text{eff}} \, (C(t) - C_{\text{eff}}(t)) \,\,\,\, \text{and} \,\,\,\, C_{\text{eff}}(0) = 0 . \nonumber 
\end{align}
where $C(t)$ and $C_{\text{eff}}(t)$ are the concentrations of drug in the 
central compartment and in the so-called effect compartment, respectively. Due to non-identifiability, the volume in both compartments is set to unity by convention here. Furthermore, $k_{e}$ 
is the elimination rate constant and $k_{\text{eff}}$ is the PK parameter 
which governs the delay between the concentration in the central compartment
and the concentration in the effect compartment. The parameter $k_{e}$ is parametrized using the elimination half-life $T_{e}$, that is $k_{e} = \frac{\text{log(2)}}{T_{e}}$. 
The parameters $k_{e}$ and 
$k_{\text{eff}}$ are assumed to be known from previous analyses, for example 
from another previously studied indication or pre-clinical data.

TITE-PK uses an adapted EWOC criterion. For this purpose, the measure of the interest is the probability of a patient experiencing at least one DLT within the first cycle (shortly the end-of-cycle 1 DLT probability), $P(T \leq t^{*}|d,f)$, where $d$ and $f$ refer to the dose and frequency of administration, respectively. Using basic event history analysis \cite{Kalbfleisch2002}, we have the following equation
\begin{align}
P(T \leq t^{*}|d,f) = 1 - e^{-H(t^{*}|d,f)}, \label{eq:Hazard}
\end{align}
which describes the relationship between the end-of-cycle 1 probabilities and the cumulative hazard function $H(t)$. All patients without a DLT up to the end of cycle 1 will be censored at the end of cycle 1, and patients with a DLT are censored at the time of a DLT. Using Equation~\eqref{eq:Hazard}, it can be shown that 
\begin{align}
\text{cloglog}(P(T \leq t^{*} | d, f)) = \log(\beta) + \log(\text{AUC}_{E}(t^* | d, f)) \label{eq:beta}
\end{align}
where $\text{cloglog}(x) = \text{log}(-\text{log}(1-x))$ and $\text{AUC}_{E}(t)$ is the area under the curve of the exposure
measure over time.

To help prior specification, $E(t)$ is obtained by scaling $C_{\text{eff}}(t)$ using a reference schedule (reference dose $d^{*}$ and frequency $f^{*}$) at the end of the first treatment cycle (cycle 1: $t^{*}$) such that
\begin{align}
\text{AUC}_{E}(t^{*}|d^{*},f^{*}) = 1.  \label{eq:AUC}
\end{align}

By combining Equation~\eqref{eq:beta} and Equation~\eqref{eq:AUC}, it follows that for the reference schedule $\text{cloglog}(P(T \leq t^{*} | d^*, f^*)) = \log(\beta)$, which we use for the prior specification
of the $\beta$ parameter. This relationship suggest to constrain $\beta$ to be positive, which ensures that
$h(t) \geq 0$, since $E(t) \geq 0$ for all $t$ (see Equation~\eqref{eq:exposure}).

The posterior distributions of end-of-cycle 1 DLT probabilities are classified into three categories in order to inform dose-escalation decisions:
\begin{enumerate}
  \item[(i)]\hspace{1cm}    $P(T \leq t^{*}|d,f)< 0.20$ \hspace{0.55cm} Underdosing (UD)
  \item[(ii)]   $0.20 \leq P(T \leq t^{*}|d,f) \leq 0.40$ \hspace{0.5cm} Targeted toxicity (TT)
  \item[(iii)]\hspace{1cm}  $P(T \leq t^{*}|d,f) > 0.40$ \hspace{0.55cm} Overdosing (OD)
\end{enumerate}

The EWOC criterion is fulfilled, if the overdosing probability $P(P(T \leq t^{*}|d,f) > 0.40)$ is smaller than the feasibility bound $a$. As the feasibility bound, we use 0.25, which is recommended by Babb et al \cite{SIM:SIM793}. Analogous to the monotonicity of dose-DLT probability assumption of CRM, TITE-PK assumes the monotonicity of the exposure measure and the end-of-cycle 1 
DLT probability. That is, $\text{AUC}_{E}(t^{*}|d,f)$ is proportional to the end-of-cycle 
1 DLT probabilities. 

In the case of sequential investigation of multiple schedules, initially TITE-PK is used to conduct the phase I trial with $S_{1}$ until the MTD is declared or trial is stopped since all doses are found to be too toxic. In this step, the frequency of administration is the same for dose-escalation decisions. Then, cohorts are recruited with Schedule $S_{2}$. 
For dose-escalation decisions, the information from the phase I trial with $S_{1}$ is treated as data together with the new information generated from the phase I trial with $S_{2}$. Since TITE-PK is an exposure-response model, there is no need to re-scale the doses from different schedules to make them comparable. As opposed to BLRM MAP and B-CRM methods, data from the completed trials is treated as part of the data instead of as part of the prior distribution.


\subsection{Software implementation}
We implemented TITE-PK in \texttt{Stan} \cite{JSSv076i01} via \textbf{rstan} R package, which employs a modern Markov chain Monte Carlo engine. For the application and simulations, four parallel chains of 1,000 MCMC iterations after warm-up of 1,000 iterations are generated. Convergence diagnostics are checked
using the Gelman-Rubin statistics and traceplots in the application. There were no
divergences reported for the implementation of the application.
The R and Stan code to analyze the everolimus application is publicly available from Github (\href{https://github.com/gunhanb/TITEPK_sequential}{https://github.com/gunhanb/TITEPK\_sequential}). The main programming code is the Stan code from the linked folder, which conducts the Bayesian computation to calculate posterior distributions. The method can be applied by changing R-code based on the application, for example different doses or schedules, while keeping the Stan code.

\section{Results}
\subsection{Simulation study} 
\begin{table}
\centering
\caption{Scenarios 1-6 in the simulation study. Doses with dose limiting toxicities in the targeted toxicity interval (0.20 - 0.40) are in boldface. Scenarios 1-6 represent phase I trials with one schedule, that is daily schedule.}
\label{t1:sc1}
\begin{tabular}{lllllll}
  \toprule
             &        \multicolumn{6}{c}{\textbf{Doses in mg/$\text{m}^2$}} \\
             \midrule
  Scenario    & 2.5 & 5 & 7.5 & 10 & 12.5 & 15\\ 
  \midrule
  1    & 0.05 & 0.10 & \textbf{0.20} & \textbf{0.30} & 0.50 & 0.70    \\ 
  2    & \textbf{0.30} & \textbf{0.40} & 0.52 & 0.61 & 0.76 & 0.87    \\
  3    & 0.05 & 0.06 & 0.08 & 0.11 & 0.19 & \textbf{0.34}   \\
  4    & 0.06 & 0.08 & 0.12 & 0.18 & \textbf{0.40} & 0.71  \\
  5    & 0.10 & \textbf{0.22} & \textbf{0.31} & 0.45 & 0.60 & 0.72 \\
  6    & 0.50 & 0.55 & 0.61 & 0.69 & 0.76 & 0.87 \\
                                       
   \bottomrule
\end{tabular}
\end{table}

We compared the operating characteristics of TITE-PK and alternative methods in a simulation study. The simulation study follows the clinical scenario evaluation framework
introduced by Benda et al \cite{bendaCSE} and it is 
inspired by the everolimus trial. Firstly, we considered scenarios only involving one schedule to compare the performance of TITE-PK to CRM and BLRM. For the CRM implementation, we used a one-parameter power model via the R package \texttt{bcrm} \cite{Bcrm2019}. Both TITE-PK and BLRM recommends the highest dose among the doses which satisfy the EWOC criteria, while CRM recommends the dose which has a DLT probability closest  to the target probability. TITE-PK and CRM have one parameter, while BLRM has two parameters in the model. Daily doses 
of 2.5, 5, 7.5, 10, 12.5, and 15 (mg/$\text{m}^2$) are investigated. The starting dose is 2.5 mg/$\text{m}^2$ for all methods. Scenarios 1-6 are summarized in Table~\ref{t1:sc1}. Doses within the targeted toxicity intervals (0.20 - 0.40) are varied based on the scenarios. Scenario 6 is an extreme scenario, where all doses are in the overdosing interval. 

 \begin{table}
\centering
\caption{Scenarios 7-13 in the simulation study. Daily doses with dose limiting toxicities in the targeted toxicity interval (0.20 - 0.40) are in boldface.}
\label{t3:sc2}
\begin{tabular}{llllllllllllll}
  \toprule
             &    &    \multicolumn{6}{c}{\textbf{Doses with Schedule $S_{1}$}} 
             &         \multicolumn{6}{c}{\textbf{Doses with Schedule $S_{2}$}} \\
             \cmidrule{3-8}  \cmidrule{9-14}
  Scenario & Schedule & 2.5 & 5 & 7.5 & 10 & 12.5 & 15  & 2.5 & 5 & 7.5 & 10 & 12.5 & 15\\ 
  \midrule
  7  & $S_{1}$ & 0.05 & 0.07 & 0.09 &  0.10 & 0.13 & 0.18 & &  &  & & &  \\ 
     & $S_{2}$ &  &  &  &  &  &  & 0.08 & 0.12 & 0.16 & 0.18 & \textbf{0.23} & \textbf{0.27}    \\
  8  & $S_{1}$ & 0.08 & 0.12 & 0.16 & \textbf{0.20} &  \textbf{0.23} &  \textbf{0.27} & &  &  & & &  \\ 
     & $S_{2}$ &  &   &  &  &  &  & 0.18 & \textbf{0.26} & \textbf{0.34} & 0.45 & 0.49 & 0.55 \\
  9  & $S_{1}$ & 0.03 &0.12 & \textbf{0.28} & \textbf{0.40} & 0.54 & 0.62 & &  &  & & &  \\ 
     & $S_{2}$ &  &  &  &  &  &  &  \textbf{0.20} & \textbf{0.30} & 0.45 & 0.50 & 0.60 & 0.75   \\
  10 & $S_{1}$ &  0.10 & \textbf{0.20} & \textbf{0.34} & \textbf{0.40} & 0.49 & 0.55 & &  &  & & &  \\ 
     & $S_{2}$ &  &  &  &  &  &  &   \textbf{0.35} &  \textbf{0.40} & 0.45 & 0.57 & 0.67 & 0.80  \\
  11 & $S_{1}$ &  0.05 & 0.07 & 0.09 & 0.15 & \textbf{0.22} & \textbf{0.28} & &  &  & & &  \\ 
     & $S_{2}$ &  &  &  &  &  &  & \textbf{0.30} & \textbf{0.35} & 0.48 & 0.52 & 0.61 & 0.70   \\
  12 & $S_{1}$ & 0.45 & 0.50 & 0.55 & 0.65 & 0.75 & 0.85 & &  &  & & &  \\ 
     & $S_{2}$ &  &  &  &  &  &  & 0.48 & 0.56 & 0.62 & 0.70 & 0.80 & 0.88   \\
  13 & $S_{1}$ & 0.18 & \textbf{0.26} & \textbf{0.34} & 0.45 & 0.49 & 0.55 & &  &  & & &  \\ 
     & $S_{2}$ &  &  &  &  &  &   & 0.08 & 0.12 & 0.16 & 0.18 & \textbf{0.23} & \textbf{0.27}   \\

   \bottomrule
\end{tabular}
\end{table}
 
We also consider Scenarios 7-13 representing sequential 
phase I trials with two schedules. In the first step, doses of 2.5, 5, 7.5, 10, 12.5, 15 (mg/$\text{m}^2$) with the dosing frequency of 48 hours ($S_{1}$) and in the second step, doses of 2.5, 5, 7.5, 10, 12.5, 15 (mg/$\text{m}^2$) with daily dosing ($S_{2}$) are administered.
The starting dose for Schedule $S_{1}$ is 2.5 mg/$\text{m}^2$. For Schedule $S_{2}$, the MTD declared for $S_{1}$ is used as the starting dose. Scenarios 7-13 are summarized in Table~\ref{t3:sc2} and displayed in Figure 1.
Doses from Schedule $S_{2}$ with DLT probabilities within the targeted toxicity intervals (0.20 - 0.40) and discrepancy between DLT probabilities of two schedules are varied based on the scenarios. Scenario 11 is a scenario in which the discrepancy of dose-toxicity curve between the schedules is higher than other scenarios. All doses are in the overdosing interval in Scenario 12. Scenario 13 is Scenario 8 with DLT probabilities for Schedules 1 and 2 switched.
Hence, the monotonicity assumption of the exposure and DLT probabilities is violated in Scenario 13. In other words, for the same dose, toxicity is higher with the lower frequent administration. Note that the weekly doses from everolimus are not chosen in the simulations in order to better investigate the monotonicity assumption of DLT probability and exposure. This is because, with the described doses and schedules in the simulations, we can easily vary the order of DLT probability of the same dose with different schedules in the scenarios.

We consider three methods for sequential phase I trial scenarios (Scenarios 7-13): TITE-PK, Bridging CRM (B-CRM), BLRM using MAP prior (BLRM MAP). As explained in the introduction, a sequential phase I trial consists of two steps. In B-CRM, the first step is conducted using the CRM, whereas BLRM is used for the first step of BLRM MAP. In B-CRM, multiple skeletons are constructed using the data from Schedule $S_{1}$. The Bayesian model averaging is used to estimate toxicity probabilities with multiple skeletons and to inform the dose-escalation decisions. We used the publicly available R-code which is provided as the supplementary material of Liu et al \cite{Liu2015}. Dose skipping is not allowed in B-CRM. We refer to Liu et al \cite{Liu2015} for more details of B-CRM. In BLRM MAP, a meta-analytic-predictive (MAP) prior is created based on the data from Schedule $S_{1}$. The MAP prior is used to construct the prior for the parameters of the BLRM. BLRM MAP uses the EWOC criterion to avoid to impose more patients to the overly toxic doses. For BRLM and BLRM MAP, the feasibility bound of 0.25 is used, as recommended in Neuenschwander et al \cite{SIM:SIM3230}. Full description of BLRM MAP is given in Neuenschwander et al \cite{doi:10.1080/19466315.2016.1174149}.
In TITE-PK and BLRM MAP, dose-escalation by more than 100\% mg/$\text{m}^2$ is not allowed. The R package \texttt{OncoBayes2} \cite{OncoBayes} can be used to implement BLRM MAP.

For TITE-PK, we need to determine PK parameters. By mimicking the everolimus trial, PK parameters are chosen as follows. The 
elimination rate constant is taken as $k_{e} = \frac{\text{log}(2)}{30}$ (1/h). 
For $k_{\text{eff}}$, an estimate is derived using the cycle length and the absorption
rate. Specifically, a log-normal distribution is constructed by matching the inverse
of cycle length 1/504 (1/h) and the absorption rate 2.5 (1/h) as the 0.025 and 0.975
quantiles, respectively. This gives a log-normal distribution with mean parameter 
0.37, hence we assume that log($k_{\text{eff}}$) = 0.37.

Prior skeletons and distributions are constructed so that prior DLT probabilities from different methods are similar. For TITE-PK model, reference dose and reference dosing frequency are determined using 7.5 
mg/$\text{m}^2$ ($d^*= 7.5$ mg/$\text{m}^2$) and 24 hours ($f^*= 1/24$ 1/h). A normal weakly informative prior (WIP) is chosen such
that log($\beta$) $\sim$
$\mathcal{N}(\text{cloglog}(P(T \leq t^{*} | d^{*}, f^{*}) = 0.30),
1.25^2)$. This implies that prior median of DLT probability at the reference dose and frequency is 0.30.
For BLRM MAP, we choose a WIP assuming median DLT probability of 0.30 at dose 7.5 mg/kg. More specifically, we choose a bivariate normal distribution
$(\log(\alpha_{1}), \log(\alpha_{2})) \sim \text{BVN}(\mathbf{m},
\Sigma)$
with means $m_{1} = \text{logit}(0.30)$ and $m_{2} = 0$, standard
deviations $\sigma_{1} = 2$ and $\sigma_{2} = 1$, and correlation $\rho =
0$. The target probability for the CRM is 0.30, that is the midpoint of the targeted toxicity interval (0.20 - 0.40).
For the CRM, the prior skeleton is calculated using the method of Lee and Cheung \cite{Lee2009} assuming an indifference interval of 0.10, which produces (0.02, 0.12, 0.30, 0.50, 0.68, 0.80). A normal prior with mean 0 and standard deviation 2 is used as the prior for the power parameter $\alpha$ in the CRM and B-CRM ($\alpha \sim \mathcal{N}(0, 2^2)$), as suggested by Liu et al \cite{Liu2015}.

The following simulation settings and decision rules are used for TITE-PK, BLRM and BLRM MAP. The maximum number of patients per trial was set to 60. If all doses are in the overdosing interval based on the EWOC criterion, the trial is stopped without selecting any dose as the MTD. Otherwise, the trial continues until the recommendation of the MTD. The recommended MTD must meet the following conditions:
\begin{enumerate}
\item[(i)] At least 6 patients have been treated at the MTD.
\item[(ii)] A minimum of 21 patients have already been treated in the trial.
\end{enumerate}

For the CRM and B-CRM, the trial is terminated for safety, if the following rule is satisfied: $P(\pi_{1} > 0.30) < 0.90$ where $\pi_{1}$ is the DLT probability of the lowest dose. The sample size of 21 patients is used unless the trial is stopped due to the safety. For all methods in the simulations, cohort sizes of 3 are used and data for 1,000 trials were generated per scenario.

\subsection{Simulation results} 

\begin{table}
\centering
\caption{Simulation results for TITE-PK, CRM, and BLRM in Scenarios 1-6.}
\label{t2:res1}
\begin{tabular}{lllllll}
  \toprule
            &   \multicolumn{6}{c}{\textbf{Scenario}} \\ 
  \midrule
           & 1 & 2 & 3 & 4 & 5 & 6\\
  \midrule  
            \multicolumn{7}{c}{Probability of selecting MTD in the targeted toxicity interval} \vspace{0.2cm}  \\
  TITE-PK  & 0.78 & 0.52 & 0.75 & 0.36 & 0.71 & n/a           \\
  CRM      & 0.73 & 0.61 & 0.24 & 0.22 & 0.79 & n/a           \\
  BLRM     & 0.75 & 0.49 & 0.64 & 0.14 & 0.78 & n/a  \\

            \multicolumn{7}{c}{Probability of selecting MTD in the overdosing interval} \vspace{0.2cm}  \\
  TITE-PK  & 0.11 & 0.03 & n/a & 0.06 & 0.17 & 0.11          \\
  CRM      & 0.09 & 0.04 & n/a & 0.04 & 0.10 & 0.14            \\
  BLRM     & 0.06 & 0.02 & n/a & 0.04 & 0.10 & 0.07   \\
            \multicolumn{7}{c}{Probability of selecting no combination as MTD} \vspace{0.2cm}    \\
  TITE-PK  & 0.01 & 0.42 & 0.00 & 0.01 & 0.04 & 0.87        \\
  CRM      & 0.01 & 0.36 & 0.01 & 0.01 & 0.03 & 0.86            \\
  BLRM     & 0.01 & 0.48 & 0.01 & 0.01 & 0.04 & 0.92  \\
            \multicolumn{7}{c}{Mean number of patients enrolled} \vspace{0.2cm}                   \\
  TITE-PK  & 24.7 & 15.4 & 23.3 & 27.0 & 22.8 & 8.1        \\
  CRM      & 20.9 & 15.7 & 20.9 & 20.8 & 20.5 & 8.9          \\
  BLRM     & 23.6 & 14.9 & 24.2 & 24.8 & 21.9 & 7.3  \\
              \multicolumn{7}{c}{Proportion of patients enrolled in the overdosing interval} \vspace{0.2cm}                   \\
  TITE-PK  & 0.28 & 0.15 & n/a & 0.13 & 0.27 & 1.00        \\
  CRM      & 0.05 & 0.05 & n/a & 0.01 & 0.06 & 1.00          \\
  BLRM     & 0.10 & 0.08 & n/a & 0.11 & 0.11 & 1.00   \\
            \multicolumn{7}{c}{Proportion of DLT observed} \vspace{0.2cm}                        \\
  TITE-PK  & 0.28 & 0.38 & 0.21 & 0.25 & 0.30 & 0.52             \\
  CRM      & 0.18 & 0.33 & 0.11 & 0.15 & 0.22 & 0.51                   \\
  BLRM     & 0.21 & 0.35 & 0.15 & 0.20 & 0.24 & 0.50 \\
   \bottomrule
\end{tabular}
\end{table}

The simulation results for Scenarios 1-6 are summarized in Table~\ref{t2:res1}. We calculated six different metrics to evaluate the performance of different methods. Scenarios 1-6 represent phase I trials with one schedule investigated. In Scenario 1, TITE-PK slightly outperforms other methods in terms of recommending the MTD in the targeted toxicity interval. The corresponding percentages are 78\% for TITE-PK, 75\% for BLRM and 73\% for CRM. Also, BLRM yields slightly lower percentage for the MTD selection in the overdosing interval compared to TITE-PK and CRM. BLRM selects the MTD in the overdosing interval in 6\% of the time, while TITE-PK and CRM do this in 11\% and 9\% of the time, respectively. In Scenario 2, CRM yields higher percentage for the MTD selection in the targeted toxicity interval compared to the TITE-PK and BLRM. CRM recommends the MTD in the targeted toxicity interval in 61\% of the time, while TITE-PK and BLRM do this in 52\% and 49\% of the time, respectively. Three methods perform similarly in terms of recommending the MTD in the overdosing interval. In scenario 3, TITE-PK results in the best performance in terms of the MTD selection in the targeted toxicity interval. TITE-PK recommends the MTD in the targeted toxicity interval 75\% of the time, while BLRM and CRM do this in 64\% and 24\% of the time, respectively.

In scenario 4, all methods perform poorly in terms of selecting the MTD in the targeted toxicity, while TITE-PK results in the best performance. TITE-PK yields 36\% percentage for the MTD selection in the targeted toxicity interval, while CRM and BLRM yields 22\% and 14\%, respectively. In scenario 5, CRM (79\%) and BLRM (78\%) produces slightly higher percentages than TITE-PK (71\%) in terms of the selecting MTD in the targeted toxicity interval. In scenario 6, all doses are in the overdosing interval. BLRM (92\%) stops the trial with slightly higher percentages compared to CRM (86\%) and TITE-PK (87\%).

In Scenarios 1, 3, 4 and 5, TITE-PK and BLRM enrolls slightly higher number of patients and results in slightly higher proportions of DLT observed in comparison to CRM. Overall, none of the methods shows superior performance in terms of the investigated metrics. The results depend on the scenarios. Similar results from the comparison of BLRM and CRM was also obtained by the simulation studies in Neuenschwander et al \cite{SIM:SIM3230}.

\begin{table}
\centering
\caption{Simulation results for TITE-PK, B-CRM, and BLRM-MAP in Scenarios 7-13.}
\label{t3:res2}
\begin{tabular}{llllllllll}
  \toprule
            &   \multicolumn{9}{c}{\textbf{Scenario}} \\ 
  \midrule
           & 7 & 8 & 9 & 10 & 11 & 12 & 13 \\
  \midrule  
            \multicolumn{10}{c}{Probability of selecting MTD in the targeted toxicity interval} \vspace{0.2cm}  \\
  TITE-PK      & 0.90 &0.70 &0.94 &0.84 &0.62 &n/a &0.17 \\
  B-CRM        & 0.83 &0.50 &0.64 &0.60 &0.52 &n/a &0.77 \\
  BLRM MAP     & 0.95 &0.56 &0.77 &0.68 &0.46 &n/a &0.55 \\
            \multicolumn{10}{c}{Probability of selecting MTD in the overdosing interval} \vspace{0.2cm}  \\
  TITE-PK      & n/a &0.22 &0.05 &0.02 &0.37 &0.02 &n/a \\
  B-CRM        & n/a &0.38 &0.08 &0.00 &0.28 &0.25 &n/a \\
  BLRM MAP     & n/a &0.40 &0.21 &0.10 &0.41 &0.03 &n/a \\
            \multicolumn{10}{c}{Probability of selecting no combination as MTD} \vspace{0.2cm}    \\
  TITE-PK      & 0.00 &0.02 &0.01 &0.14 &0.00 &0.98 &0.15 \\
  B-CRM        & 0.00 &0.02 &0.02 &0.28 &0.20 &0.75 &0.00 \\
  BLRM MAP     & 0.00 &0.02 &0.02 &0.22 &0.12 &0.97 &0.01 \\
            \multicolumn{10}{c}{Mean number of patients enrolled} \vspace{0.2cm}                   \\
  TITE-PK      & 21.7 &21.7 &21.4 &19.4 &21.8  &3.7 &19.7 \\
  B-CRM        & 21.0 &21.0 &21.0 &18.0 &19.0  &9.0 &21.1 \\
  BLRM MAP     & 21.5 &23.6 &21.6 &20.0 &22.8  &4.8 &23.4 \\
                \multicolumn{10}{c}{Proportion of patients enrolled in the overdosing interval} \vspace{0.2cm}    \\
  TITE-PK      & n/a  &0.39  &0.17  &0.12  &0.61 &1.00   &n/a \\
  B-CRM        & n/a  &0.46  &0.15  &0.06  &0.72 &1.00   &n/a \\
  BLRM MAP     & n/a  &0.59  &0.40  &0.26  &0.70 &1.00   &n/a \\
            \multicolumn{10}{c}{Mean number of of DLT observed} \vspace{0.2cm}                        \\
  TITE-PK      & 5.3  &8.2  &6.2  &7.5 &10.2 &1.8  &2.4 \\
  B-CRM        & 4.5  &7.0  &7.3  &7.5 &8.5  &4.0  &3.0 \\
  BLRM MAP     & 5.7  &9.7  &7.7  &8.2 &11.1 &2.4  &3.9 \\
   \bottomrule
\end{tabular}
\end{table}

We continue with Scenarios 7-13 in which sequential phase I trials are investigated. The simulation results under Scenarios 7-13 are summarized in Table~\ref{t3:res2}. In Scenario 7, BLRM MAP produces the best performance in terms of the MTD selection in the targeted toxicity interval, while TITE-PK is the second. The corresponding percentages are 95\%, 90\%, and 83\% for BLRM MAP, TITE-PK, and B-CRM respectively. In Scenarios 8-11, TITE-PK demonstrates superior performance in terms of selecting the MTD in the targeted toxicity interval. TITE-PK selects the MTD in the targeted toxicity interval in 14\%, 17\%, 16\%, and 10\% more simulated trials in comparison to the second best performed method in Scenarios 8-11, respectively. In Scenarios 8 and 9, TITE-PK produces lower percentages in terms of the MTD selection in the overdosing interval, selecting MTD in 16\% and 3\% less simulated trials compared to BLRM MAP. In Scenario 11, CRM (28\%) displays superior performance in terms of the MTD selection in the overdosing interval in comparison to other methods. In Scenario 12, TITE-PK and BLRM MAP displays better performance than B-CRM by stopping the trial in 98\% and 97\% of the time, while requiring less patients than other methods. The monotonicity assumption of the exposure and DLT probabilities is violated in Scenario 13.
In Scenario 13, B-CRM outperforms other methods by selecting MTD in the targeted toxicity interval in 22\% more trials compared to the BLRM MAP. TITE-PK (17\%) displays the worst performance in terms of the MTD selection in the targeted toxicity interval.

In Scenarios 7-13 except 12, different methods enrolls similar number of patients. In Scenarios 7-13 except 12, in terms of the proportion of DLT observed, all methods perform similarly. In Scenarios 7-12, TITE-PK displays the best or the second best performance in terms of the MTD selection in the targeted toxicity and overdosing intervals. However, TITE-PK clearly shows poor performance in Scenario 13, which is expected, as the monotonicity assumption between exposure and DLT probability is violated.

\subsection{Revisiting the everolimus trial} 
Returning to the data set described before,
consider the everolimus trial shown in Table~\ref{t01:RawDataRAD}. Firstly, we
analyse the data only from the daily schedule using the BLRM, the CRM, and the TITE-PK. Secondly, 
we analyse it as if the trial is conducted 
sequentially, specifically $S_{1}$ is weekly schedule and $S_{2}$ is daily schedule using BLRM MAP, B-CRM, and TITE-PK. The reference schedule is determined using dosing amount of 
5 mg/$\text{m}^2$ ($d^* = 5$ mg/$\text{m}^2$) and dosing 
frequency of 24 hours ($f^* = 1/24$ 1/h). For TITE-PK, PK parameters are chosen such that
$T_{e} = 30$ (hours) and $\text{log}(k_{\text{eff}}) = 0.37$ as explained in the simulation study.

\begin{figure}[htb]
  \includegraphics[scale=0.125]{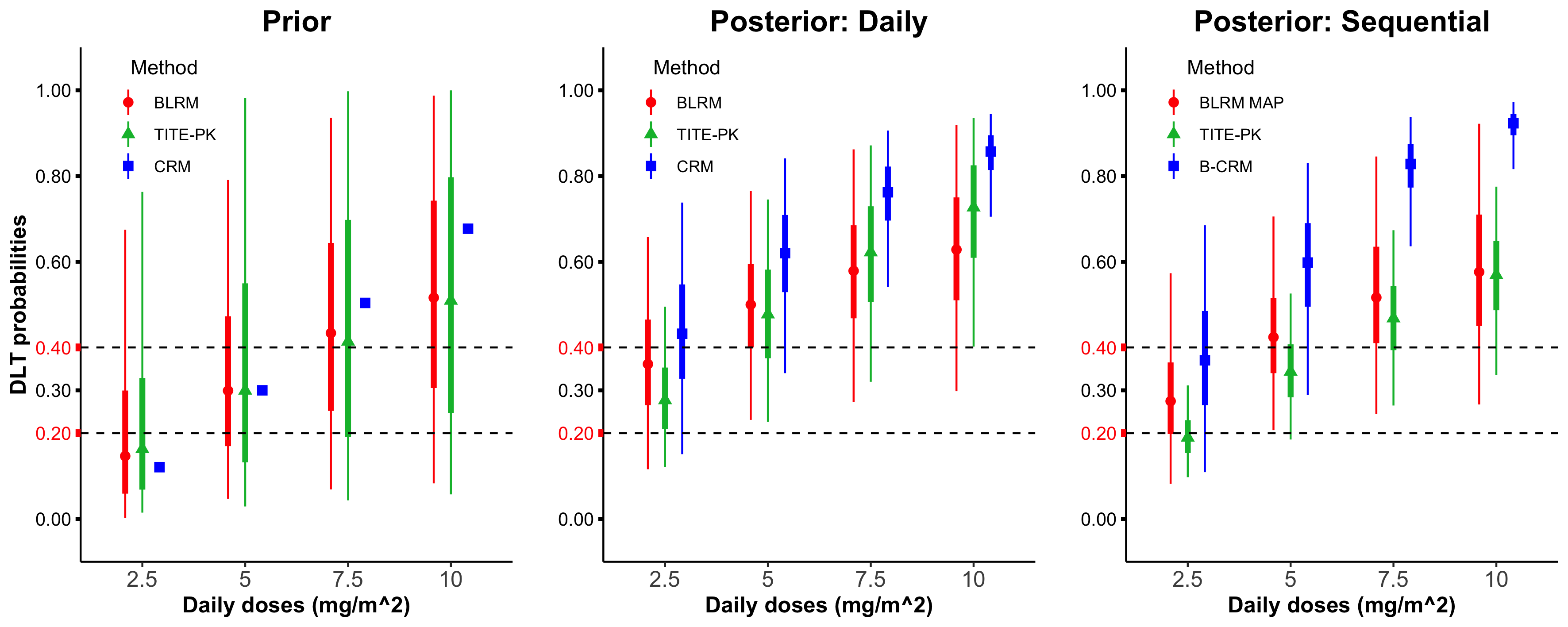}
  \caption{Everolimus trial: Prior medians (A), posterior medians daily (B), 
    and sequential (C), 50\%
    equi-tailed credible intervals (thick lines), and 95\% equi-tailed
    credible intervals (thin lines) of daily doses for DLT
    probabilities obtained by BLRM (BLRM-MAP for Sequential), CRM (B-CRM for Sequential), and for end-of-cycle 1 DLT
    probabilities obtained by TITE-PK. Prior skeletons are shown for CRM in the plot A.
     ``Sequential'' refers that 
    analysis is done by assuming the trial is conducted sequentially, namely firstly weekly schedule, secondly daily schedule. Also, ``Daily'' means data only from daily schedule
    is considered. 
    Vertical dashed lines ($0.20$-$0.40$) are the
    boundaries of the targeted toxicity interval.}
  \label{fig:RADdailyPosteriors}
\end{figure}

To compare BLRM, CRM and TITE-PK models, priors are constructed so that
prior DLT probabilities are similar. To define a WIP for
BLRM, we choose a bivariate normal prior with following parameters
$(m_{1} = \text{logit}(\pi_{d^{*}} = 0.30), m_{2} = 0, \sigma_{1} = 1.25,
\sigma_{2} = 1, \rho = 0)$. For the CRM, we use the target probability of 0.30.
The prior skeleton is, then, calculated assuming an indifference interval of 0.10, which produces (0.12, 0.30, 0.50, 0.68).
For TITE-PK, a normal WIP is chosen such that
$\log(\beta) \sim \mathcal{N}(\text{cloglog}(P(T \leq t^{*} | d^{*},
f^{*}) = 0.30), 1.25^2)$. The
summaries of prior DLT probabilities of BLRM and TITE-PK, and prior skeletons of CRM are shown in
Figure 2A. Points, thick lines and thin
lines correspond to median estimates, the 50\% and the 95\% equi-tailed
credible intervals, respectively. Vertical dashed lines (0.20-0.40)
are the boundaries of the targeted toxicity interval. Recall that, in TITE-PK and BLRM, eligible doses are determined based on the EWOC criterion, whereas CRM selects the dose closest to the target probability.

Figure 2B displays the posterior estimates of DLT probabilities, when we only consider daily schedule data. BLRM suggests that all doses are in the overdosing interval, meaning that the trial should be stopped without any dose declared as
the MTD. The estimated overdosing probability of 2.5 mg/$\text{m}^2$ is 0.40, which is higher than 0.25. 
For TITE-PK, only 2.5 mg/$\text{m}^2$ is not in the overdosing interval. The overdosing probability of 2.5 mg/$\text{m}^2$ is 0.14, $P(P(T \leq t^{*}|d = 2.5,f = 24) > 0.40) = 0.14$, which is smaller than 0.25. Although median DLT probability estimate of CRM is higher than the median DLT probability estimate of BLRM, CRM does not conclude that the trial
 should be stopped. This is because, $P(\pi_{1} > 0.30) = 0.80$, which is smaller than 0.90. Furthermore, credible intervals obtained by the CRM is getting shorter with the increasing dose, which was also observed by Neuenschwander et al \cite{SIM:SIM3230}. Overall, high overdosing probabilities for all doses seem reasonable, since 2 DLT were observed in the 4 patients with 2.5 mg/$\text{m}^2$, and 3 DLT were in the 6 patients with 5 mg/$\text{m}^2$ dose.

We continue by treating the data from the weekly schedule as the completed trial in a sequential phase I trial. We estimate the DLT probabilities of daily doses, but also taking into
consideration the data coming from the weekly data. 
To implement BLRM-MAP, the MAP prior is calculated based on the weekly data. Later, the BLRM is fitted
and posterior estimates of DLT probabilities are obtained. In the B-CRM, prior skeletons are calculated using the weekly data. Then, CRM via a Bayesian model averaging method is used to estimate DLT probabilities.
TITE-PK, naturally, combines information from
different schedules. Figure 2C displays the estimated
posterior summaries of DLT probabilities of daily doses obtained by TITE-PK, BLRM-MAP and B-CRM approaches. For both TITE-PK and BLRM-MAP, the overdosing probability of dose 2.5 mg/$\text{m}^2$ is decreased substantially, namely
from 0.40 to 0.18 for BLRM-MAP, and from 0.14 to 0.00 for TITE-PK. For CRM, the probability $P(\pi_{1} > 0.30)$ is also decreased from 0.80 to 0.67.
The reduction of the overdosing probabilities of 2.5 mg/$\text{m}^2$ seems
reasonable, since in the weekly schedule data, no DLT were observed in the 5
patients with 20 mg/$\text{m}^2$ and 4 DLT were in the 13 patients with 30
mg/$\text{m}^2$. The interval estimates of 2.5 mg/$\text{m}^2$ and 5 mg/$\text{m}^2$ obtained by TITE-PK are shorter,
hence more precise estimates compared to BLRM-MAP and B-CRM. All three methods suggest that daily 2.5 mg/$\text{m}^2$ is sufficiently safe, hence it can be declared as the MTD
which was the conclusion of the original phase I trial.

\begin{figure}[htb]
  \includegraphics[scale=0.52]{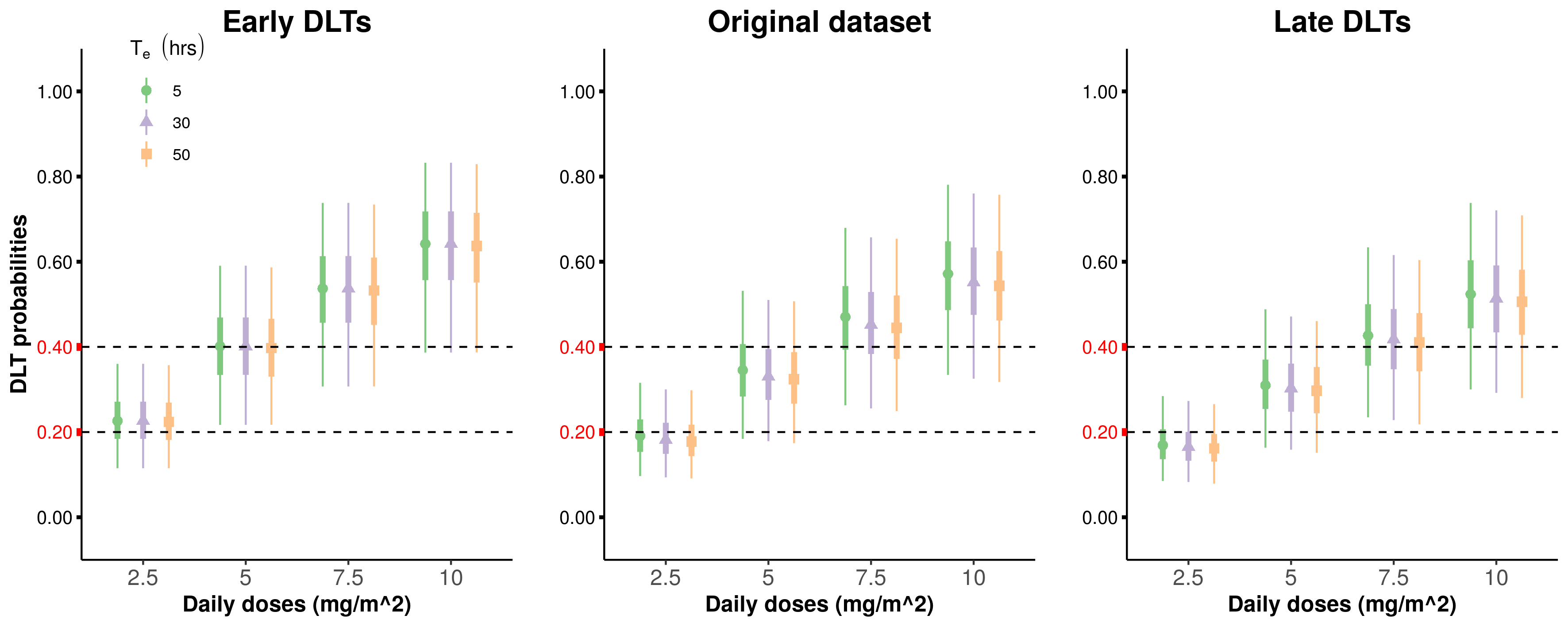}
  \caption{Misspecification of elimination half-life $T_{e}$ and
    different timing of DLT. Using different values of $T_{e}$,
    posterior median, 50\% and 95\% equi-tailed credible intervals for
    end-of-cycle 1 DLT probabilities obtained by TITE-PK for two
    hypothetical datasets (early DLT and late DLT) and the original
    everolimus trial dataset are shown. Early DLT dataset and late
    DLT dataset are created by changing timing of DLT from day 15 to
    day 1.5 and to day 20.5, respectively. Data from both weekly and
    daily schedules are included in the analysis.}
  \label{fig:RADtemisspecified}
\end{figure}

As pointed out in Methods Section, by construction of TITE-PK, the elimination
half-life $T_{e}$ is treated as known. To investigate the influence of
misspecification of the $T_{e}$ parameter, we fit TITE-PK using
$T_{e}$ ranging from 5 to 50 hours. The timing of all DLT (in total 9
DLT) were reported at day 15. To examine what would be the influence
of the timing of DLT, we also fit TITE-PK to two hypothetical
datasets. Early DLT dataset and late DLT dataset are created by
changing timing of DLT from day 15 to day 1.5 and to day 20.5,
respectively. Posterior estimates of DLT probabilities for different
$T_{e}$ values and for different timing of DLT are shown in
Figure 3. The middle plot corresponds to the
original everolimus trial data. Firstly, the posterior medians and
credible intervals obtained by different $T_{e}$ values look very
similar. In practice, a reliable estimate of elimination half-life is often not available. 
Hence, these results are reassuring for the practicality of TITE-PK. Secondly,
timing of DLT has a crucial affect on the posterior estimates, and
hence the overdosing probabilities. Having the same number of DLT,
the earlier the DLT happened, the higher the overdosing probability of
the corresponding dose estimated. This makes sense, since one would
expect the drug to be more toxic if DLT happened earlier than later.

\section{Discussion} 
In this manuscript, we have adapted TITE-PK for efficiently estimating the maximum tolerable dose in sequential phase I trials involving multiple schedules. To integrate data from different schedules, TITE-PK makes use of exposure-response modelling considering kinetic drug properties. Moreover, we have demonstrated that TITE-PK can be used as an alternative to the standard methods like the BLRM or CRM to conduct phase I trials with only one schedule. In these trials, we have demonstrated that TITE-PK displays similar performance compared to CRM and BLRM. 
In scenarios with sequential phase I trials, TITE-PK mostly displays superior performance in terms of acceptable dose recommendations in comparison to the bridging CRM and BLRM using MAP approach. An application involving weekly and daily schedules is used to illustrate TITE-PK. Also, using the application, we have shown that TITE-PK is robust against the misspecification of the PK parameter elimination half-life.

Here, we considered a sequential trial in which trial with schedule $S_{1}$ is already competed. Another type of a sequential trial can be designed to use the so-called concurrent co-data \cite {doi:10.1080/19466315.2016.1174149}. That is, the trial with Schedule $S_{1}$ is still ongoing, and we would like to utilize the information from the Schedule $S_{1}$ to inform dose-escalation decisions with Schedule $S_{2}$ (and vice versa). TITE-PK can be used for such designs as well. We did not investigate these situations, since these are beyond the scope of the paper.

In a sequential phase I trial, strata sometimes refer to other than schedules, e.g. patient populations. 
In such situations, the integration of different strata can be achieved using a MAP approach. Since TITE-PK is parametrized by mimicking the interpretable parameters of the BLRM, it can be extended to use a MAP approach like the BLRM. A key strength of the TITE-PK approach is its ability to integrate the data from different treatment schedules in a model based approach. This makes ad-hoc approaches like dose re-scaling obsolete which reduces the need for strong discounting of historical data from different schedules. However, discounting may still be needed to account for other sources like different patient populations. Recently, Li and Yuan \cite{li2020} introduced a method to find the MTD for paediatric dose-escalation trial by incorporating information from the concurrent adult data.  Their method is based on the CRM and uses Bayesian model averaging to control discounting from the adult data. The BLRM MAP approach makes the assumption of the exchangeability between different schedules. Instead of using a MAP prior, one can use exchangeability/non-exchangeability (EX-NEX) \cite{EXNEX, OncoBayes} approach for phase I trials with multiple schedules, which relaxes the exchangeability assumption.

The monotonicity assumption of the exposure and DLT probabilities is often very reasonable but could be considered a limitation of TITE-PK. Similarly, the BLRM and the CRM assumes the monotonicity of the doses and DLT probabilities.
Since, we have used a linear PK model within TITE-PK, the monotonicity of the exposure and DLT probabilities implies the monotonicity of the dose and DLT probabilities. In the simulations where we investigated phase I trials with one schedules (Scenarios 1-6), we assumed the monotonicity of dose and DLT probabilities. When there is a heavy violation of the assumption of the monotonicity (as in Scenarios 13), the operating characteristics are expected to be weaker compared to bridging CRM or BLRM MAP. The violation of the assumptions occurred, since there is a clear conflict in exposure and DLT profiles between different schedules. Such violations can be informed using the external PK data from the ongoing trial. An extension combining TITE-PK with MAP could be more useful for such situations.


\section*{Conflict of interest}
S.W. and A.S. own Novartis stakes and are employees of Novartis. T.F. is a consultant to Novartis and has served on data monitoring committees for Novartis.
Novartis is the manufacturer of everolimus, an everolimus trial was used 
to motivate and illustrate the investigations presented here (see Section 1.1 and 3.3).

\vfill

\bibliography{bibliography}%

\end{document}